\documentstyle{article}

\begin{document}

\large
\begin{flushright}
ITP.SB-93-33 \\
\end{flushright}

\begin{center}
\Large
{\bf Exact Dynamics of Quantum Dissipative  System in Constant
External Field }\\
\vspace{1cm}
\Large

Chang-Pu Sun $^\clubsuit$ and Li-Hua Yu $^\spadesuit$\\

$^\clubsuit$ Institute for Theoretical Physics, State University of New York,
Stony Brook,
NY 11794-3840\footnote{\large Department of Physics,
Northeast Normal University,Changchun 130024, P.R.China}\\
$ ^\spadesuit$725C, National Synchrotron Light Source, Brookhaven
National Laboratory, N.Y.1197
\vspace{0.4cm}

Abstract\\
\end{center}

The quantum dynamics of a simplest dissipative system, a particle
moving in  a constant external field , is exactly studied by  taking into
account its interaction with a bath of Ohmic spectral density.
We apply the main idea
and methods developed in our recent work [1] to
quantum dissipative system with constant external field.
Quantizing
the dissipative system
we obtain the simple and exact solutions for the  coordinate operator of the
system in Heisenberg picture and  the
wave function of the composite system of system and bath in Schroedinger
picture. An effective Hamiltonian for the dissipative system
is explicitly derived from these solutions with  Heisenberg picture method and
thereby the  meaning
of the wavefunction governed by it is  clarified
by analyzing the  effect of the Brownian motion. Especially, the general
effective Hamiltonian for the case with arbitrary potential is  directly
derived with this method for the case when the Brownian motion can be
ignored.
Using this effective
Hamiltonian, we show an interesting fact that the dissipation suppresses
the wave packet spreading.

\newpage
{\bf 1.Introduction}

\vspace{0.4cm}
This paper is mainly devoted to the application and generalization of
the idea and methods developed in
our recent work [1]
to a dissipative system with constant external field.
Though most discussions in this paper are proceeded based on ref.[1],
the context is written in
a self-contained  form. In ref.[1], we work on the case of harmonic
oscillator moving  in a bath with the Ohmic spectral density presented
by  Caldeira-Leggett [2].
It has been shown there
that the wavfunction of system plus bath is precisely
described by a direct product of  two independent Hilbert space, one
of which is described by an effective Hamiltonian while  the other
represents the effect of the bath, i.e., the Brownian motion.
Therefore,  this study clarifies the structure of the wavefunction of
the system dissipated by its interaction with the bath and
thereby  reveals the relationship between the different approaches for
quantum dissipative systems before. No path integral technology is
needed in this treatment. Notice that  the study of dissipative
quantum systems(DQS), especially for the
damped harmonic quantum oscillator(DHQO),
has a rather long history [3-18] and has been paid much
attention more recently due to the work by Caldeira and Leggett [2].

Now let us briefly describe two main approaches for quantum dissipative
systems before the work in ref.[1].
To reproduce  and quantize the phenomenological dissipative  equation
$$M\ddot q(t) = -\eta \dot q (t) - \frac{\partial V(q)}{\partial q},
\eqno{(1.1)}$$
for one-dimensional dissipative system S with coordinate q,  mass M
and potential V(q),
one approach is to embed it into an environment, a bath B of N harmonic
oscillators interacting with the system  S through certain coupling [3-12].
Since the bath and system constitute a closed composite system C equals to
S plus B , the quantization of C
is naturally direct. Then, its corresponding Heisenberg equation results
in the above phenomenological dissipative  equation in the operator form
by eliminating the variables of the bath through proper approximations,
such as the Markovian approximation and the Wigner-Wisskopf approximation [9].
This approach provides the motion equation with both the friction force
in dissipation process and the  fluctuation force in Brownian
motion. The path integral technique [2,13] and field theory method [14] were
used in this approach sometimes.
Along this direction, the work of  Caldeira and Leggett reveals a remarkable
fact that the dissipation can occur exactly, instead of approximately,
if the spectral density of the bath is Ohmic (to be described later). It
enlightens people to reconsider many problems of dissipation  in an exact
version.

Another approach, for DHQO, is the use of an effective Hamiltonian [15-17]
$$H_E=H_E(t)=\frac{1}{2M} e^{-\eta t/M}p^2 +{1\over 2}M\omega ^2e^{\eta t/M}
q^2.
\eqno{(1.2)}$$
which is now called Caldirora-Kani (CK)  Hamiltonian [13]. With the canonical
commutation relation
$$[q , p] = i\hbar,\eqno{(1.3)}$$
this  Hamiltonian automatically yields the dissipation equation (1.1)
through the Heisenberg equation. Notice that the alternative forms of the
effective Hamiltonian have been by many authors [17] and an elegant example can
be found in ref.[18].  Though this approach is very convenient
to treat some  dynamical problems of dissipation process for
both classical and quantum cases, such as tunnelling and motion of wave packet
, it is only a purely phenomenological method without
the microscopic mechanism constructing $H_E$  and thus the Brownian motion
can not
be analysed by making its use. Especially, the meaning of wavefunction of
the evolution governed the CK Hamiltonian is much ambiguous.

The present paper applies the method and idea developed in reference [1]
to dissipative system with a constant external field, to establish
the connection between the above two approaches and to consider how
the interaction between the bath and
the system leads to an explicit description for the dissipative system in
terms of the effective Hamiltonian. All the discussions in this paper
is proceeded with a simplest model,
a particle moving in one dimension with constant potential
field, but the main idea and methods can be generalized for other cases.
The present work is completed in an unified framework
in accompany with the answers to the entangling
questions listed as follows:\\
{\bf i}. How to quantize the dissipative system with
the effective Hamiltonian in an exact context?\\
{\bf ii}. What is the meaning of the wavefunction governed by the effective
Hamiltonian ?\\
{\bf iii}. How to construct the propagator for the dissipative system in the
Heisenberg representation ?\\
{\bf iv }. What happens to the spreading
of the wave packet in present of dissipation ?\\
\vspace{0.5cm}

{\bf 2.Exact Classical Langevin Equation}
\vspace{0.4cm}

In this section
we reformulate the first approach about quantum dissipative system
in Ohmic case with a simplest example that  the dissipative system S is
considered
as a charged particle with unit mass, unit negative charge and coordinate
q in a constant
electric field E. It interacts with a bath B of N harmonic oscillators $B_i$ of
coordinates $x_j$, mass $m_j$ and frequency $\omega_j$. Let p and $p_j$ be
the corresponding momentums to q and   $x_j$ respectively. The Hamiltonian
of the composite system of S plus  B is written as
$$H=\frac {1}{2}p^2-Eq+\sum^{N}_{j=1}[\frac {p_j^2}{2m_j}+\frac {1}{2}
m_j \omega_j^2(x_j-q)^2]$$
$$=\frac {1}{2}p^2-Eq+H_B -\sum^{N}_{j=1}C_jx_jq +\Delta V. \eqno{(2.1)}$$
where
$$H_B = \sum^{N}_{j=1}[\frac {p_j^2}{2m_j}+\frac {1}{2}m_j \omega_j^2x_j^2]$$
The problem of renormalization of interaction for $N=\infty$ mentioned in
ref.[2] is naturally enjoyed by the frequency-dependent coupling constant $C_j=
m_j \omega_j^2$ and  the renormalized potential
$$\Delta V= \sum^{N}_{j=1}\frac {C_j^2}{2m_j \omega_j^2}q^2 =
 \sum^{N}_{j=1}\frac {1}{2}m_j \omega_j^2 q^2$$
 in the Hamiltonian (2.1).

Through the canonical equation, the Hamiltonian (2.1) defines a system of
classical motion equations
$$\ddot{q}=E-\sum^{N}_{j=1}C_jx_j-\sum^{N}_{j=1}\frac {C_j^2}{m_j \omega_j^2}q
,\eqno{(2.2a)}$$
$$\ddot{x_j}=-\omega_j^2x_j-\frac {C_j}{m_j}q.\eqno{(2.2b)}$$
Notice that the canonical equation for the closed composite system C of
S plus B determines  the ordinary momentum-velocity relations
$$  p(t) = \dot q(t),~~~~  p_j(t)=m_j\dot x_j(t) .  $$
 By making use of  the Laplace transformations of eqs.(2.2), a direct
 of substitution of eq.(2.2b) into eq.(2.2a) yields an exact  motion
equation for the system S
$$\ddot{q}=E+L(q)+G(t),\eqno{(2.3)}$$
where
$$G(t)=\sum^{N}_{j=1}C_j[x_j(0)cos \omega_jt+\frac{\dot{x_j}(0)}{\omega_j}
sin\omega_jt],\eqno{(2.4)}$$
and the second term in the right hand side in eq.(2.3)
$$L(s)=\wp^{-1}[-s^2\sum^{N}_{j=1} \frac {C_j^2 s^2}{m_j\omega_j^2(s^2+
\omega_j^2 )}\bar{q}(s
)] \eqno{(2.5)}$$
is determined by the inverse of Laplace transformation $\wp$
$$\wp[q(t)]=\bar{q}(s) =\int ^{\infty}_0q(t)e^{-st} dt$$

Usually, for the finite N or the general spectral distribution $\rho (\omega_j)
$ of infinite oscillators in bath, the dissipative term $-\eta \dot q$ for
a positive number $\eta $ does not
appear exactly. However, according to Caldeira and Leggett,
a specific  spectral distribution of the bath
$$\rho (\omega_j)=\frac{2\eta \omega_j^2m_j}{\pi C_j^2}=
\frac{2\eta}{\pi m_j\omega_j^2},\eqno{(2.6)}
$$
which is called Ohmic distribution , enable the sum over index j
$$\sum^{N}_{j=1}\frac {C_j^2 s^2}{m_j\omega_j^2(s^2+\omega_j^2 )}$$
to become an  integral
$$\int ^{\infty}_0\rho (\omega_j)\frac {C_j^2
s^2}{m_j\omega_j^2(s^2+\omega_j^2)}
d\omega_j
=\frac{2\eta s^2}{\pi}\int ^{\infty}_0\frac {d\omega_j}{(\omega_j^2+s^2) }
=s\eta.$$
and thus  results in
 $$L(q)=-\eta \dot q-\eta q(0)\delta(t)$$
immediately through  an inverse Laplace transformation. Then,
the dissipative equation - the classical Langevin equation
$$\ddot q=E-\eta \dot q +G(t),\eqno{(2.7)}$$
occurs as an exact evolution based on the elementary  Hamiltonian dynamics.
Because the dissipative process is invertable, it is significant to pay our
attention only to the process with $t > 0$ as
follows. In this sense the impact $\delta(t)$ in the
dissipative equation does not play role in  dynamical problems.

Notice that the above Ohmic distribution (2.6) is only an alternative,
but explicit
and convenient reformulation of the Caldeira-Leggett's constraint
$$J(\omega)=\eta\omega$$
on spectral density
$$J(\omega)=\frac {\pi}{2}\sum^{N}_{j=1}
\frac {C_j^2 }{m_j\omega_j }\delta(\omega-\omega_j)
.\eqno{(2.8)}$$

It is also remarked that the fluctuating external force G(t) acting on
system S is due to the effect of the bath and depends on the initial
states of the oscillators in the bath.
For  the classical statistical thermal average $<~~>_{classical}$, G(t) obeys
the dissipation-fluctuation relation at temperature T
$$<G(t)G(t')>_{classical}=  \frac {1}{2}\eta KT\delta(t-t') \eqno{(2.9)} $$
and $<G(t)>_{classical}=0$. It actually is the classical Brownian force in
dissipation process.
In concluding
this section it is pointed out that the  classical Langevin equation
has been studied for a long time by many authors by certain approximations
, but here our emphasis  is that it can  appear  exactly with Ohmic
spectral density for the bath.
\vspace {0.4cm}

{\bf 3.Quantization of Dissipation with Effective Hamiltonian}
\vspace {0.4cm}

In the following sections, we first detail the description in ref.[1] of the
the system plus bath with  the present example. Then, we
develope this theory to derive a general effective Hamiltonian for
arbitrary potential.

It can be observed from eq.(2.7) that, in Ohmic case, the action of bath B on
the system S can be exactly described as two parts, the dissipative term
$-\eta \dot q$ only depending on the state of S  and the Brownian force G(t)
depending on the initial state of bath. In order to study various dynamical
problems in dissipative
process, it is necessary to  determine in what sense the dissipative system can
be isolated  from the environment as a `quasi-closed' system only depending on
its own  variables. Such a `quasi-closed' system  can evolve
independently and the effect
of the bath is totally enjoyed by the friction coefficient $\eta$.
For the classical case, it is quite clear that at zero temperature the
Brownian force can be neglected for the vanishing dissipation-fluctuation
relation. According to the fluctuation-dissipation relation eq.(2.9),
the classical statistical fluctuation of the bath    is proportional to
$\eta T$ . Only when T
= 0,   the system is isolated with an
effective dissipation equation
$$\ddot q=E-\eta\dot q,\eqno{(3.1)}$$
and the whole effect of bath on the system is only characterized  through
the  friction
constant $\eta$. However, for the quantum case, the story is not so direct.
Now, we  consider the composite
system C equals S plus B. Since C is closed its  quantization process is well
build
in elementary quantum mechanics.

Let us start with the exact solution to eqs.(2.2b) and (2.7) ,
$$q(t)=Q(t)+X(t):$$
$$Q(t)=a(t)\dot q(0)+q(0)+g(t)$$
$$X(t)= \sum^{N}_{j=1}X_j(t):\eqno{(3.2)}$$
$$X_j(t)=\alpha_k(t)x_k(0)+\beta_k(t)\dot x_k(0),$$
where
$$ a(t)=t_\eta={1-e^{-\eta t}\over \eta},~~~~ g(t)={E\over\eta}(t-t_\eta)$$
$$\alpha_k(t)= \frac{C_j}{\eta^2+\omega_k^2}[\frac{\eta}{\omega_k}
sin [\omega_kt]-\frac{\omega_k}{\eta}(cos[\eta t]-1)-\eta a(t)]$$
$$\beta_k(t)= \frac{C_j}{(\eta^2+\omega_k^2)\omega_k}[\frac{\eta}{\omega_k}
(cos[\omega_kt]-1)-\frac{\omega_k}{\eta}sin[\eta t] + \omega_k a(t)].$$
The  canonical commutation relations at t = 0 for the closed  system of S plus
 B
$$[q(0),\dot q(0)]=i\hbar,~~~[x_j(0),\dot x_j(0)]=\frac
{i\hbar}{m_j} \eqno{(3.3)}$$
quantize q(t) and $\dot q(t)$ at any instant t so that
$$[q(t),p(t)]=i\hbar,~~~[x_j(t), p_j(t)]=i\hbar \eqno{(3.4)}$$
for the ordinary momentum-velocity relations
 $$p(t) = \dot q(t), p_j(t)=m_j\dot x_j(t).$$
Notice that the quantized coordinate q(t) is separated into two commuting
part Q(t) and X(t) depending on the system and bath respectively.

Now, we consider the quantum statistical problem of the dissipation
and fluctuation  for the above linear system. The quantum
statistical average $<~~~>=<~~~~>_{quantum}$  over the bath is defined by
$$<A>=\frac{Tr[A e^{-\beta H_B}]}{Tr[e^{-\beta H_B}]},~~~~~~~~
\beta=\frac{1}{KT}$$
for an observable A. Here, $H_B$
is the Hamiltonian for bath. A direct calculation gives the quantum fluctuation
- dissipation relation
$$<G(t)>=0,$$
$$D(t-t')_{T\rightarrow 0}=\frac{1}{2}<\{G(t),G(t')\}>=\frac{\eta\hbar}{\pi}
\int_{0}^{\infty}\omega_jcoth(\frac{\beta \hbar\omega_j}{2}) cos[\omega_j(t-t')
]d\omega_j.\eqno{(3.4)}$$

At high temperature limit, ie., $\beta
\rightarrow 0$ or $T  \rightarrow \infty$
, the above results approach the classical dissipation-fluctuation
relation (2.9 ). In zero temperature limit, it becomes

$$D(t-t')_{T\rightarrow 0}=\frac{\eta\hbar}{\pi}\int_{0}^{\infty}\omega_j
cos[\omega_j(t-t')]d\omega_j$$
$$=\frac{\eta\hbar}{\pi}{\Large lim}_{\mu \rightarrow \infty}\{\mu^2[\frac{sin
[\mu(t-t')]}{\mu(t-t')}-
{1\over 2}\frac{sin^2[{\mu(t-t')\over 2}]}{({\mu(t-t')\over 2})^2}]\}
\eqno{(3.5)}$$
This equation concludes that the Brownian force can hardly be neglected
even at zero
temperature for quantum case,
but the corresponding fluctuation is  proportional  to  $\eta \hbar$. So this
is a quantum fluctuation.
The effective coordinate Q(t) can be used
to approach  the physical coordinate q(t) only when
this quantum fluctuation is neglected in certain sense. In next section
we will clarify further the exact meaning of this argument in terms of the
wavefunction description.

When the quantum fluctuation can be  ignored, the evolution  of the
dissipative system may
be approached by the variable Q(t) independent of the bath. Then,  the
system is  isolated from bath to produce a ``closed '' dynamics. Now,
let us derive the effective Hamiltonian governing  this kind of dynamics. Our
derivation is valid for both the classical and quantum case

By making an observation that the explicit  expression (3.2) for Q(t)
determines the
commutator
$$[Q(t),\dot Q(t)]=i\hbar e^{-\eta t},\eqno{(3.6)}$$
it is quite natural to define the canonical momentum-velocity relation
$$P(t)=e^{\eta t}\dot Q(t)\eqno{(3.7)}$$
for the effective variable Q so that  the basic commutators for the Hamiltonian
dynamics is
$$[Q(t),P(t)]=i\hbar.\eqno{(3.8)}$$
Notice that the definition of the canonical momentum based on eq.(3.6)
is not unique and the different definitions such as in ref.[18] may give
different forms of the effective Hamiltonian.
For the given Q(t) changing as eq.(3.2), one can explicitly obtain
expressions of $\dot Q(t)$ and $\dot P(t)$ in terms of Q(t) and P(t)
$$\dot Q(t)=-Ee^{\eta t}Q(t),~~~~\dot P(t)= e^{-\eta t}P(t),\eqno{(3.9)}$$
Associated with Heisenberg equation
$$\dot Q(t)=\frac{1}{i\hbar}[Q(t),H_e(t)],$$
$$\dot P(t)=\frac{1}{i\hbar}[P(t),H_e(t)],$$
the eqs.(3.9) lead to a simple system of partial differential equations
about the effective Hamiltonian $H_e(t)$
$$\frac{\partial H_e(t)}{\partial Q }=-Ee^{\eta t}Q,~~~\frac{\partial
H_e(t)}{\partial P}= e^{-\eta t}P(t). \eqno{(3.10)}$$
A solution to eq.(3.10)
determines  an effective Hamiltonian $H_e(t)$
$$H_e=H_e(t)=\frac{1}{2} e^{-\eta t}P^2-EQe^{\eta t}+F(t).\eqno{(3.11)}$$
up to an arbitrary function F(t)  independent of P and Q.
It can be regarded a generalization of
CK  Hamiltonian [14-16] of damped
harmonic oscillator.

In spirt of the above discussions, for an arbitrary
potential V(Q), one guesses the general effective Hamiltonian for  dissipation
problem
$$H_E=H_E(t)=\frac{1}{2} e^{-\eta t}P^2+V(Q)e^{\eta t}.\eqno{(3.12)}$$
Indeed, the Heisenberg equation of $H_E=H_E(t)$ with the basic commutator
(3.8) can result in the dissipation equation
$$\ddot Q(t)=-\eta \dot Q(t)-\frac{\partial V(Q)}{\partial Q}.\eqno{(3.13)}$$
for arbitrary potential.

In fact, the generalized CK Hamiltonian (3.12) can be directly  derived
in the explicit equations satisfied  by  Q(t) and P(t) as Heisenberg equation
. It follows from eq.(3.13) that
the commutator $[Q(t),\dot Q(t)]$ at time t satisfies an equation
$${d\over dt}[Q(t),\dot Q(t)]= -\eta  [Q(t),\dot Q(t)].\eqno{(3.14)}$$
Even in present of an arbitrary potential V(q), it still leads to
the same commutator
$$[Q(t),\dot Q(t)]=i\hbar e^{-\eta t}.\eqno{(3.6')}$$
as that for case of harmonic oscillator  and the case of  constant
external field.
Then, eq.(3.6)   suggests the same canonical momentum-velocity relation
$$P(t)=e^{\eta t}\dot Q(t)\eqno{(3.7')}$$
as eq.(3.7) for the two above mentioned cases. Using the
dissipative equation  eq.(3.13) and eq.(3.7'), we have
$$\dot P(t)=- e^{-\eta t}{\partial V(Q)\over \partial Q} \eqno{(3.15)}$$
Equations (3.15, 3.7') and the Heisenberg equation determine the equations of
commutators about unknown-Hamiltonian $H_E$
$$[ Q(t),H_E(t)]= i\hbar e^{\eta t}P(t)/M,\eqno{(3.16)}$$
$$[ P(t),H_E(t)]=-i\hbar e^{-\eta t}{\partial V(Q)\over \partial Q},$$
Obviously, the generalized CK Hamiltonian (3.12) is just a solution of above
equations. Therefore, we derive out the
generalized CK Hamiltonian for the arbitrary potential V(q).
Notice that the derivation of the generalized CK Hamiltonians (3.12)
and its special case (3.11) here can also make sense
for the classical case without Brownian motion so long
as one use the commutators yinsteaded of the Poisson brackets.

\vspace{0.5cm}

{\bf 4.Meaning of Wave Function for Dissipation}

In the last section we derived the effective Hamiltonian in terms of the
Heisenberg equations. It can serve as a starting point studying some dynamical
problems of quantum dissipation under certain  sense.
To understand the
problem completely, one must investigate the physical meaning of
the (effective) wavefunction defined by the effective Hamiltonian (3.11)
and thereby clarify in what sense this effective Hamiltonian
can be used correctly.
Thus, we turn to the Schroedinger picture and
detail the main ideas and methods developed in ref. [1].

Let us first show that the wavefunction of the
composite system C equals S plus B can be reduced to a direct
product  of  wavefunctions in in two independent Hilber spaces.
To this end, it
is observed that the coordinate operator
$$q(t)=Q(t)+X(t)$$
of S is the sum of two commuting parts Q(t) and X(t). So the  eigenfunction
of q(t)  with eigenvalue  q is expressed as a direct product
$$|q,t>=|Q,t>\otimes |\xi_1 ,t>\otimes |\xi_2 ,t>\otimes...\otimes  |\xi_N ,t>
\eqno{(4.1)}$$
of the the eigenstates $|Q,t>$ of Q(t) and $|\xi_j ,t>$ of $X_j(t)$ with the
eigenvalues Q and $\xi_j$ respectively. Notice that these eigenvalues satisfy
$$q=Q+\sum ^{N}_{j=0}\xi_j.\eqno{(4.2)}$$
Due to eq.(4.2), for a given q, there exist many different sets of
$\{\xi_1,\xi_2,...,\xi_N\}$
corresponding to  q , that is to say, these eigenstates $|q,t>$
are degenerate. So, the new notation $|q,\{\xi_j\},t>$ with additional index
$\{\xi_j\}$
is needed, instead of $|q,t>$,  to distinguish among the different degenerate
eigenstate with the same eigenvalues q. Correspondingly, the  following
notation for arbitrary complex number $\xi_j$
$$ |Q,\{\xi_j\}>=|Q>\otimes |\xi_1 >\otimes |\xi_2 >\otimes...\otimes|\xi_N >
\eqno{(4.3)}$$
is used to represent  the degenerate eigenstate of q(0)
with eigenvalue Q , which belong to  the Hilbert space
$$V=V_S\otimes V_B=V_S\otimes\prod^N_{j=1}\otimes V_j$$
of the composite system C. Here, $V_S$ and $V_B=\prod^N_{j=1}\otimes V_j$
is the Hillbert spaces of S and B respectively; $V_j$ is the Hillbert
space  for the j'th oscillator in the bath B; $|Q>$
is the eigenstates with eigenvalue Q and  $|\xi_j> (\in V_j )$
the eigenstates of  $x_j(0)$ with  eigenvalue $ \xi_j $.

Let the composite system be initially in a product state
 $$|\psi(0)>=|\phi>\otimes|W>=|\phi>\otimes \prod^N_{j=1}\otimes
|W_j>,\eqno{(4.4)}$$
at t=0
where $|\phi>$, $|W>$ and $|W_i>$ belong to $V_S$,  $V_B$ and
$V_j$ respectively.
The central problem we should face is whether
the product form similar to eq.(4.4), where the wavefunction
is a direct product of two wavefunctions in  two independent Hilbert spaces,
is persevered in the Schrodinger evolution.
The positive answer will implies that the system can be isolated from the bath
by making use of an effective Hamiltonian. To consider this problem,
it is necessary to  calculate the
evolution operator U(t) or its matrix elements. Because the coordinate operator
 q(t), for general value of t,  can formally be regarded as
an unitary transformation
$$q(t)=U(t)^\dagger q(0)  U(t)$$
of q(0), the eigenstate
$$|q,\{\xi_j\},t>=U(t)^\dagger |q,\{\xi_j\}>\eqno{(4.5)}$$
of q(t) with eigenvalue q can be constructed in terms of the evolution operator
and  the eigenstate $|Q=q,\{\xi_j\}>$ of q(0) with the same eigenvalue q.
Then, the  coordinate component of the evolution state
$$|\psi(t)>=U(t)|\psi(0)> $$
can be calculated as
$$\Psi(Q,\{\xi_j\})=<Q,\{\xi_j\}| \psi(t)>=<Q,\{\xi_j\}|U(t)|\psi(0)>$$
$$=
[<\psi(0)|U(t)^\dagger|Q,\{\xi_j\}>]^*=[<\psi(0)|Q,\{\xi_j\},t>]^*\eqno{(4.6)}$$
according to the eigenstate $|Q,\{\xi_j\},t>$ of q(t) for general t. In fact,
this eigenstate  can be directly solved in the coordinate representation with
$$q(0)=Q,\dot q(0)=-i\hbar \frac{\partial }{\partial Q},x_j(0)=\xi_j,
\dot {x_j(0)}=-i\hbar /m_j\frac{\partial }{\partial \xi_j }$$
Obviously, the eigenstates $|Q,t>$ of Q(t) and $|\xi_j,t>$ of $X_j(t)$
with eigenvalues $xi_j$ :
$$\phi_Q(Q',t)=<Q'|Q,t>=e^{\frac{i}{\hbar a(t)}[-\frac{1}{2}Q'^2+(Q-g(t))Q'
+\lambda(Q)]}$$

$$u_{\xi_j}(\xi_j',t)=<\xi_j'|\xi_j,t>=e^{\frac{i}{\hbar \alpha _j(t)}
[-\frac{1}{2}\beta_j(t)\xi_j'^2+\xi_j\xi_j'+\mu(\xi_j)] }\eqno{(4.7)}$$
are explicitly obtained from the differential equations
$$[-i\hbar a(t)\frac{\partial }{\partial Q'}+Q'+g(t)]\phi_Q(Q',t)=
Q\phi_Q(Q',t)
$$
$$[-i\hbar \alpha _j(t)\frac{\partial }{\partial \xi_j'}+\beta_j(t)]
u_{\xi_j}(\xi_j',t)=\xi_ju_{\xi_j}(\xi_j',t).\eqno{(4.8)}$$
where $\lambda(Q)$ and $\mu(\xi_j)$ are the functions independent
of $Q'$ and $\xi_j'$ respectively.

Now, we can write
$$<Q',\{\xi_j'\}|q,\{\xi_j\},t> =<Q'|Q=q-{\tiny \sum^N_{j=1}}
\xi_j,t>\prod^N_{j=1}<\xi_j'
|\xi_j,t>$$
$$=\phi_{q-\sum \xi_j}(Q',t)\prod^N_{j=1}u_{\xi_j}(x_j',t),\eqno{(4.9)}
$$
which results in
$$\Psi(q,\{\xi_j'\})=[<\psi(0)|q,\{\xi_j'\},t>]^*
=W(q-\sum \xi_j,t)^*\prod^N_{j=1}W_j(\xi_j,t)^*\eqno{(4.10)}$$
where
$$
W(q-\sum \xi_j)=\int dQ'<\phi(0)|Q'>\phi_{q-\sum \xi_j}(Q',t)$$
$$W_j(\xi_j,t)=\int d\xi_j'<w_j|\xi_j'>u_{\xi_j}(\xi_j',t)$$

Notice that the variables $\xi_j$  are related to the bath, but they
are not the coordinates $x_j$ of bath. Therefore,  applying the above
analysis to
a practical problem, one should distinguish between $\xi_j$ and $x_j$.

It is observed from eq.(4.10) that , if the variable q is so highly
excited that  the Brownian motion contribution $\sum \xi_j$ is small enough
in comparison with q in certain sense, the first factor in the right hand side
of eq.(4.10) is approximately independent of $\xi_j$. In this sense, the whole
wavefunction for composite
system C is factorized in to two independent parts  belong to $V_S$ and $V_B$
respectively. The first part $W$ represents
the wave function of the dissipative
system isolated to evolve according to the effective Hamiltonian (3.11).
Because of
the  Brownian motion, the physical variable $q=Q+\sum\xi_j$ fluctuates about
Q through
the  mean value  of
$(\sum \xi_j)^2$
$$<(\sum \xi_j(t))^2> =
\sum^{N}_{j=0}\frac{\hbar}{2m_j\omega_j}(|\alpha_j(t)|^2
+\omega_j^2|\beta_j(t)|^2)coth\frac{\hbar\omega_j}{2KT}
\eqno{(4.11)}$$
at temperature T. It is zero at t = 0 and approach its final value
$$<[\sum \xi_j(t=\infty)]^2> =\frac{\hbar}{2m_j\omega_j}(\frac{\pi}{2}
+arctg [\frac{\omega_0}{\eta \omega}]).\eqno{(4.12)}$$
with low temperature limit in the time of the order of $1/\eta$. Obviously,
whether the effective Hamiltonian can work well or not mainly depends on
whether the values $<(\sum \xi_j(t))^2> $ can be  neglected in practical
problems or not.

 For the further study in the dynamics of the dissipative system, we need
to calculate its propagator. There are usually two ways to do that
in principle:
1). By solving the effective Schrodinger equation
$$i\hbar \frac{\partial}{\partial t} |\Psi(t)> =H_E(t)|\Psi(t)>,
$$
directly for the obtained effective Hamiltonian $H_E$.
2). By using reduced density matrix  in terms of the path integral for
the composite system C equals S plus B.
Now, we will deal with this problem along a shorter road , as the third way
in  which we need not to know what is the effective Hamiltonian $H_E$.
Having  known the time-evolution of the observable Q(t) explicitly, we
can  derive the propagator from Q(t) directly without use of the effective
Hamiltonian $H_E$.
$H_E$ will thereby be derived in a  purely quantum mechanical
 way. This approach not only avoids the complexity in calculations in
the first  two  approaches,  but also give us new insight for the
understanding of quantum dissipation.

According to the definition of propagator, we have
$$G(q_2,\{\xi_{j,2}\} t_2,q_1,\{\xi_{j,1}\},t_1)=
<q_2,\{\xi_{j,2}\} ,t_2|q_1,\{\xi_{j,1}\},t_1>$$
$$=<Q_2,t_2|Q_1,t_1>\prod_{j=1}^{N}<\xi_{j,2},t_2|\xi_{j,1},t_1, >$$
$$=G(Q_2,t_2,Q_1,t_1)\prod_{j=1}^{N}G_j(\xi_{j,2},t_2|\xi_{j,1}t_1).
\eqno{(4.13)}$$
Because of the linearity of of the Heisenberg equation in the variables
$Q(t),\dot Q(t)$ and $\xi_j(t),\dot \xi_j(t)$, the operators $O(t_2)
(O=Q, \dot Q ,\xi_j ,\dot\xi_j ) $ at time $t_2$  must be  a linear combination
of the operators  $O(t_1)$ at $t_1$. For example,
$$Q(t_1)=Q(t_2) +  a(t_1, t_2)P(t_2)+ b(t_1, t_2)$$
$$Q(t_2)=Q(t_1) +  a(t_2, t_1)P(t_1)+ b(t_2, t_1)\eqno{(4.14)}$$
where
$$ b(t_2, t_1) = g(t_2)-g(t_1)-Ea(t_1) e^{-\eta t_1}a(t_2,t_1) $$
$$a(t_2,t_1)=-a(t_1,t_2)=a(t_2)-a(t_1) . $$
Then, the definitions of eigenstates of $Q_i(t)$
$$Q(t_i)|Q_i,t_i>=Q_i|Q_i,t_i>,i=1,2$$
and its $Q_j, (j\neq i)$ -representation leads to the partial
differential equations for the propagator
$$[Q_2+ a(t_1,t_2).i\hbar \frac{\partial}{\partial Q_2}+ b(t_1,t_2)]
G(Q_2,t_2;Q_1,t_1) = Q_2G(Q_2,t_2;Q_1,t_1) $$
$$[Q_1- a(t_1,t_2).i\hbar \frac{\partial}{\partial Q_1}+ b(t_2,t_1)]
G(Q_2,t_2;Q_1,t_1)^* = Q_1G(Q_2,t_2;Q_1,t_1)^*\eqno{(4.15)} $$

Solving the above equations, we obtain the propagator
$$G(Q_2,t_2;Q_1,t_1) = \frac{1}{ \sqrt {\frac{2i\pi }{\eta}|e^{-\eta t_1}
-e^{-\eta t_2}|}}$$
$$\times exp[\frac{i}{a(t_1,t_2)\hbar }\{{1\over 2} Q_1^2+{1\over 2}  Q_2^2
-Q_1Q_2+ b(t_2,t_1) Q_1+b(t_1,t_2)Q_2+\theta(t)\}]\eqno{(4.16)}$$
where $\theta(t)$ is an arbitrary function of time  independent of
$Q_i,(i=1,2)$ Similarly, we can also calculate the factors
$G_j(\xi_{j,2},t_2;\xi_{j,1}t_1)$, but here we need not to explicitly
write them out for our present purpose. For highly-excited q-system where the
Brownian motion is ignored, the first
factor $ G(Q_2,t_2;Q_1,t_1) $
can be regarded as  the effective propagators for the dissipative system.
By taking into account that the propagator, in fact,  is the  certain
matrix elements of the evolution operator $U_Q(t)$, i.e.,
$$<Q_2|U_Q(t)|Q_1>=<Q_1|U_Q(t)^\dagger|Q_2>^*=<Q_2,t|Q_1>=G(Q_2,t;Q_1,0)$$
the effective Hamiltonian (3.11) can also be derived again from
$$ H_E= i\hbar\frac{\partial U_Q(t)}{\partial t}U_Q(t)^{-1}\eqno{(4.17)}$$
Notice that this derivation of the effective Hamiltonian here is {\it purely
quantum mechanical}.

\vspace{0.5cm}

{\bf 5. Spreading of Wave Packet Suppressed by Dissipation}
\vspace{0.4cm}

In this section the effective Hamiltonian (3.12) for
the dissipative system is applied
to study a quite simple dynamical problem:
the motion of the wave packet of a `free' (E=0)
particle of mass M in one dimension.
This study will show an interesting fact that
the dissipation must suppress the spreading of the wave packet
if the breadth of initial wave packet is so wide that the effect
of  Brownian motion  can be ignored. Usually , without dissipation,
a Gaussian  wave packet will infinitely spreads into the whole space and the
localization
of  wave will lose  completely in  evolution process. In future direction its
 breadth increases as t  until infinity while its height decreases from
its initial value to zero. Notice that the height and breadth of a wave packet
are correlated through its normalization. However, for the present case with
dissipation, there appears a quite different picture about the wave packet
spreading. It will be proved that the final breadth and height have the finite
limit value as $t\rightarrow \infty$. In the following we use
$M\neq 1$ and denote by $x$
the operator $Q$ approaching the physical coordinate $q$.

Starting with the effective Schrodinger equation
$$i\hbar \frac{\partial }{\partial t}\Psi(x,t)=-{\hbar^2\over 2M} e^{-\eta t/M}
\frac{\partial^2 }{\partial x^2}\Psi(x,t)\eqno{(5.1)}$$
governed by the effective Hamiltonian in absent of the external field,
the evolution of wave function is generally expressed by
$$\Psi(x,t)=\sum_k <k|\Psi(x,0)>exp[ikx -{iE_k t_\eta\over \hbar} ]
\eqno{(5.2)}$$
where
$$E_k={\hbar^2k^2\over 2M}$$
is the energy of the momentum eigenstate $|k>$:
$$<k|x>={1\over \sqrt{2\pi}}e^{ikx}.$$
Notice that,  for the definition
$$t_\eta =a(t) ={M(1-e^{-\eta t/M)}\over \eta}$$
we have  $t_\eta \rightarrow t $ when $\eta /M\rightarrow 0$. So one
can regard  $t_\eta $ as a $\eta$-deformation of time t. Especially, $t_\eta $
approaches a limit $M\over \eta$ as $t\rightarrow\infty$. This fact
will enjoys the physical features of wave packet spreading in present
of dissipation.

Take the Gaussian wave packet
$$\Psi(x,0)=<x|\Psi(0)>= {1\over [2\pi d^2]^{1\over 4}}e^{ik_0x-\frac{x^2}{4d}}
\eqno{(5.3)}$$
as an initial state, in which
$$<\Psi(0)|x|\Psi(0>= 0, <\Psi(0)|P|\Psi(0>=\hbar k_0,$$
$$<\Delta
x>=\sqrt{<\Psi(0)|x^2|\Psi(0)>-<\Psi(0)|x|\Psi(0)>^2}=d.\eqno{(5.4)}$$
The above equation (5.4) shows that the wave packet is centered at  x=0 and
has an average
momentum $\hbar k_0$. Its breadth is d.  According to the wave equation (5.1),
the Fourier transformation
$$\Psi(x,0)=\frac{1}{\sqrt{2\pi}}\int^{\infty}_{\infty}\psi(k)e^{ikx}dk$$
$$\psi(k)=\frac{2d^2}{\pi}e^{-d^2(k-k_0)^2}
$$
for the initial wave packet (5.3) determines
the wavefunction at t
$$\Psi(x,t)=\frac{exp[ik_0x -{iE_k t_\eta\over \hbar}]}{(2\pi)^{1/4}\sqrt{d
+it_\eta\hbar/(2Md)}} exp[-{1\over4}(x-{k_0t_\eta\hbar/M})^2
\frac{1-it_\eta\hbar/2Md^2}{d^2+(t_\eta\hbar/2Md)^2}]\eqno{(5.5)}$$

To understand the physical meaning represented by the wavefunction (5.5),
we write down the corresponding position probability:
$$|\Psi(x,t)|^2= \frac{1}{\sqrt{2\pi[d^2
+(t_\eta\hbar)^2/(2Md)^2]}}exp[-\frac{(x-{k_0t_\eta\hbar/M})^2}{2[d^2
+(t_\eta\hbar)^2/(2Md)^2]}].\eqno{(5.6)}$$
The above formula tells us that the motion of  the wave packet
in  dissipation
process starts  with initial velocity
$$v_0=<P/M>= \hbar k_0/M$$
and  its center is initially at the position x=0. Subjected to time
evolution,  the center of the  Gaussian wave packet will stop
at a limit position
$$x_{limit}=\frac{\hbar k_0}{\eta}.$$
as $t\rightarrow\infty$. In this process, the velocity of the center
$$v(t)=\frac{d}{dt}<\Psi(x,t)|x|\Psi(x,t)>= \frac{\hbar k_0}{M}e^{-\eta t/M} $$
decrease from $\frac{\hbar k_0}{M} $ to zero. The above fact means  that
the  motion
of the center of wave packet is as the same as the that of  a dissipative
classical particle. However, a purely quantum picture
is manifested by the finite change of its breadth
$$B(t)=\sqrt{d^2+(\hbar t_\eta/2 M d)^2}$$
from $d$ at t=0 to a limit value
$$B_{limit}=\sqrt{d^2+(\hbar/2\eta  d)^2}$$
as $t\rightarrow\infty$. Both the position at which the center of wave packet
finally stop and its final breadth are independent of the mass M! It
defines the limit shape of  wave packet
at which the spreading wave packet finally takes. These physical features
are illustrated by the figures 1 and 2.

It is finally pointed out that this  suppressing of wave packet spreading
by dissipation possibly provides a mechanism to localize a quantum particle.
\vspace{0.5cm}

{\bf Acknowledgements\\}
\vspace{0.5cm}

The authors whish to express their sincere
thanks to Professor C.N. Yang for drawing our attention to the
problem of dissipative systems, for spending his
valuable time in many sessions of
stimulating discussions on this subject, and for many suggestions
which are critically important for the ideas of this paper.
The part of work by Chang-Pu Sun is supported in part by the
Cha Chi- Ming fellowship through the CEEC program at the State
University of New York at Stony Brook, and in part by the NSF of China through
the Northeast Normal University. The part of work by Li- Hua Yu
is performed under the auspices of the U.S.
Department of Energy under Contract No. DE-AC0276CH00016.

\newpage

{\bf References}
\vspace{0.5cm}
\begin{enumerate}

\item L.H.Yu, C.P.Sun, {\it Wavefunction Evolution of a Dissipative System,},
NSLS Brookhaven National Laboratory
and SUNY.SB.ITP preprint, May 1993,submitted for publication
in Phys.Rev.Lett.
\item A.O. Caldeira, A.J. Leggett, Ann. Phys. 149, 374 (1983), and
Physica 121A, 587 (1983)
\item R.Zwanzig, J. Chem. Phys. 33, 1338 (1960)
\item M.J.Lax,J.Phys.Chem.Solid 25,487(1964);Phys.Rev.A145,110(1966)
\item P.Ullersma,Physica,32,27(1966)
\item H.Haken,Rev.Mod.Phys.47,67(1975)
\item Senitzky, Phys. Rev. 119, 670 (1960)
\item G.W.Ford,J.T.Lewis,R.F.O'comell,Phys.Rev.A.37,4419,(1988)
\item W.H. Louisell, "Quantum Statistical Properties of Radiation",
John Wiley and  Sons (1973)
\item G.W. Ford, M. Kac, P. Mazur, Jour. Math. Phys., 6, 504 (1965)
\item S. Nakajima, Prog. Theor. Phys. 20, 948 (1958)
\item M. D. Kostin, J. Chem. Phys. 57, 3589 (1972)
\item R.P. Feynman, F.L. Vernon, Ann. Phys. 24, 118 (1963)
\item K.Fujikawa,S.Iso,M.Sasaki,H.Suzuki,Phys.Rev.Lett.68,1093(1992)
\item P.Caldirola,Nuovo Cimento,18,393,(1941)
\item E. Kanai, Prog. Theor. Phys. 3, 440 (1948)
\item H. Dekker, Phys. Report, 80, 1 (1981)
\item H.W.Peng, in {\it Guangzhou Particle Physics Conference of 1980},
p.57-67, Science Press,1981.
\end{enumerate}
\vspace{1cm}

\Large\bf
Figure Captions\\
\large\bf

 Figure 1\\

The Motion of Wave Packet  in Present of Dissipation :
$F(x,t_\eta)=|\Psi(x,t)|^2$. The dissipation
not only restrict the motion of the center of the Gaussian wave packet
like a classical particle, but also  suppress its  spreading so that
it takes a limit Gaussian wave packet with finite breadth and height
as $t\rightarrow \infty$.

Figure 2\\

The projection of {\bf Figure 1} on the $|\Psi|^2-x$ Plane. It actually
represent the figures $F(x,t_\eta)=|\Psi(x,t)|^2 $
at t=0, t=T, t=2T, t=3T ,...,t=NT. When
$N\rightarrow \infty$, there is a lower Gaussian wave packet
$|\Psi(x,t=\infty)|^2=F(x,1/\eta)$.

\end{document}